\documentclass[trackchanges]{aastex7}

\begin{document}

\title{Sensitivity of spectral lines to granulation: from the Sun to K-type stars}
\shortauthors{Vasilyev et al.}

\correspondingauthor{V.~Vasilyev}
\email{vasilyev@mps.mpg.de}

\author[0009-0009-3020-3435]{V.~Vasilyev}
\affiliation{Max-Planck-Institut f\"ur Sonnensystemforschung, Justus-von-Liebig-Weg 3, 37077 G\"ottingen, Germany}
\email{}

\author[0000-0002-3243-1230]{K.~Sowmya}
\affiliation{Institute of Physics, University of Graz, Universit\"atsplatz 5, 8010 Graz, Austria}
\email{}

\author[0000-0002-8842-5403]{A.~I.~Shapiro}
\affiliation{Institute of Physics, University of Graz, Universit\"atsplatz 5, 8010 Graz, Austria}
\affiliation{Max-Planck-Institut f\"ur Sonnensystemforschung, Justus-von-Liebig-Weg 3, 37077 G\"ottingen, Germany}
\email{}

\author[0000-0002-6087-3271]{N.~Kostogryz.}
\affiliation{Max-Planck-Institut f\"ur Sonnensystemforschung, Justus-von-Liebig-Weg 3, 37077 G\"ottingen, Germany}
\email{}

\author[0000-0003-1971-5551]{D.~Vukadinović}
\affiliation{Institute of Physics, University of Graz, Universit\"atsplatz 5, 8010 Graz, Austria}
\email{}

\author[0000-0002-0929-1612]{V.~Witzke}
\affiliation{Institute of Physics, University of Graz, Universit\"atsplatz 5, 8010 Graz, Austria}
\email{}
\author[0000-0002-6568-6942]{T.~Bhatia}
\affiliation{Max-Planck-Institut f\"ur Sonnensystemforschung, Justus-von-Liebig-Weg 3, 37077 G\"ottingen, Germany}
\email{}

\author[0000-0002-8863-7828]{A.~Collier~Cameron}
\affiliation{Centre for Exoplanet Science, SUPA School of Physics and Astronomy,\\
University of St Andrews, North Haugh, St Andrews KY16 9SS, UK}
\email{}

\author[0000-0001-7696-8665]{L. Gizon}
\affiliation{Max-Planck-Institut f\"ur Sonnensystemforschung, Justus-von-Liebig-Weg 3, 37077 G\"ottingen, Germany}
\affiliation{Institut für Astrophysik und Geophysik, Georg-August-Universit\"at G\"ottingen,  37077  G\"ottingen, Germany
}
\email[]{}

\author[0000-0002-3418-8449]{S.~K.~Solanki}
\affiliation{Max-Planck-Institut f\"ur Sonnensystemforschung, Justus-von-Liebig-Weg 3, 37077 G\"ottingen, Germany}
\affiliation{School of Space Research, Kyung Hee University, Yongin, Gyeonggi 17104, Republic of Korea}
\email{}

\begin{abstract}
Stellar granulation is a source of radial-velocity (RV) jitter of $1$~m/s on Sun-like stars, limiting the detection of Earth analogs. One way to move beyond this limit is to use line-weighting schemes to mitigate granulation-driven RV variability in a spectral-type-dependent manner. A physics-based granulation-sensitivity diagnostic is essential. Here, we use a line-by-line sensitivity measure from 3D magneto-convection simulations that quantifies how strongly each spectral line's Doppler shift and strength respond to convective velocity and thermodynamic fluctuations. Extending our previous solar analysis, which used the spatial variability of line profiles across a single granulation snapshot as a computationally efficient proxy for temporal variability, we test the transferability of this diagnostic to cooler stars and characterize how line sensitivity varies with spectral type. We synthesize high-resolution spectra with the \texttt{MPS-ATLAS} code from 3D time-dependent MURaM simulations representative of the Sun and late-G and K dwarfs, focusing on neutral and singly ionized iron lines spanning a wide range of excitation potentials and strengths. As $T_{\mathrm{eff}}$ decreases, convective velocities weaken and the ionization balance shifts, producing a clearer separation between line families: Fe~{\sc i} lines show lower velocity sensitivity and smaller fractional strength variability, whereas Fe~{\sc ii} lines show the opposite. Using cumulative contribution functions, we further relate spectroscopic velocity jitter to characteristic line-formation temperature. The diagnostic continues to separate robust and sensitive lines in late-G and K dwarfs, enabling spectral-type-aware cross-correlation masks and line-by-line weights. Solar-optimized line selections are not generally portable to cooler stars, especially when based on equivalent-width stability rather than velocity sensitivity. 
\end{abstract}


\section{Introduction} \label{sec:intro}
With continued advances in instrumental stability, extreme-precision radial-velocity (EPRV) searches for Earth analogs are gradually becoming more limited by stellar "jitter" from photospheric convection and magnetic activity, rather than by the spectrograph itself. To circumvent activity-driven signals, RV surveys therefore tend to focus on the quietest, least magnetically active stars. However, even these stars exhibit significant variability from granulation (the photospheric manifestation of near-surface convection: hot, rising granules and cooler, sinking intergranular lanes), which produces time-variable spectral line profiles and Doppler shifts that obscure planetary RV signals. For example,  for a  relatively quiet star like the Sun, granulation can induce disk-integrated RV fluctuations of order $\sim 1$~m/s \citep{Lakeland2024}, far larger than the $\sim 10$~cm/s semi-amplitude expected for an Earth-mass planet in the habitable-zone orbit around a Sun-like star. Meanwhile, the instrumental precision of modern spectrographs  already reaches the $\sim10$-$30$~cm/s regime \citep{Fischer2016EPRRVSTATE}, making the dominant error budget astrophysical rather than instrumental \citep{Zhao2023, Dumusque2025}.

A promising way to mitigate astrophysical RV noise is to exploit the fact that different physical processes imprint on spectral lines in different ways. The Doppler motion induced by an orbiting planet produces a common shift of all photospheric lines, whereas granulation and magnetic activity can introduce apparent RV shifts in a line-dependent manner. In the context of granulation, this line dependence arises because different spectral lines respond unevenly to the photospheric inhomogeneities produced by convection. Their formation depths differ, so they sample different mixtures of up-flowing granules and down-flowing lanes and inherit different convective shifts and asymmetries \citep{Gray2005book, Nordlund2009}. Crucially, this granulation imprint is not set by formation depth alone: the atomic parameters of a transition (excitation potential, oscillator strength, and ionization stage) regulate how line opacity responds to local temperature and density fluctuations, thereby changing the relative contribution of granules and lanes to the emergent line profile and, consequently, the line’s apparent RV variability \citep[see, e.g.,][]{Asplund2000, Asplund2005}. Toward cooler spectral types, granulation contrasts weaken and the ionization/opacity balance shifts, changing the physical conditions in the line-forming layers \citep[see, e.g.,][]{Magic2013, Beeck2013}.   In particular, the changing balance between majority/minority species (e.g., Fe~{\sc i} versus Fe~{\sc ii}) and increasing molecular/blend opacity can relocate formation layers and modify line-by-line convective biases relative to the solar case \citep[e.g.][]{Gray2005book, Magic2013, Beeck2013}.

Traditionally, cross-correlation RV extraction techniques co-add information across thousands of lines using instrument-tuned binary masks or templates \citep{Baranne1996}. In this framework, the measured RV shift is a weighted average over an ensemble of lines \citep{Bouchy2001, Pepe2002, Anglada2012ApJS}. Because this procedure delivers one RV per exposure (for a given mask), it does not explicitly track how different subsets of lines respond to stellar variability, limiting the ability to exploit differential line sensitivities within a single-mask RV time series.

Line-by-line approaches track per-line sensitivities and then select and weight lines accordingly to mitigate (or, when desired, isolate) stellar variability-driven signals \citep{Dumusque2018, Cretignier2020, Moulla2022}. A prerequisite for such analyses is therefore a quantitative estimate of each line's sensitivity to granulation and activity, an estimate that can be informed directly by physics-based spectral synthesis from realistic convection simulations. Empirically, the payoff can be substantial: \citet{Palumbo2024} modeled time-resolved solar line variability and reported a $\sim 30$~cm/s spread between the most and least variable lines, underscoring the leverage available through line selection and sensitivity-informed weighting. 

Physics-based spectral synthesis from realistic 3D radiation-(magneto)hydrodynamic simulations provides a natural way to predict these line-by-line sensitivities from first principles  \citep{Dravins2021, Dravins2023,Sowmya2025, Frame2025}. In particular, \citet{Sowmya2025} (hereafter Paper~I) predicted the granulation-driven RV sensitivities of Fe~{\sc i} and Fe~{\sc ii} lines in the Sun by post-processing three-dimensional radiative MHD MURaM simulations \citep{Vogler2005} of near-surface solar convection with radiative-transfer calculations using the \texttt{MPS-ATLAS} code. This provides a basis for constructing line masks or line-weighting schemes for EPRV analyses in the solar context \citep{Witzke2021}. 

Here we extend the same framework from the Sun to cooler late-G and K dwarfs. K dwarfs are among the most favorable stellar hosts for EPRV searches for habitable-zone terrestrial planets. K dwarfs are especially attractive targets because they have long main-sequence lifetimes and typically show weaker magnetic activity and fewer energetic flares than cooler M dwarfs \citep{Cuntz2016, Richey2019ApJ}. 
From an RV standpoint, K dwarfs also yield larger semi-amplitudes than G dwarfs for planets of the same mass, because their habitable zones lie at shorter orbital periods and their stellar masses are lower.  Toward cooler spectral types the physical conditions at the depths where lines form change, thereby modifying the line-by-line responses to convection relative to the solar case. We again focus on  Fe~{\sc i} and Fe~{\sc ii} lines, because they are abundant in the spectra of the Sun and late-G/K stars.  We quantify how granulation sensitivity varies across spectral type, relate these trends to changes in line-formation depth, and discuss the implications for line selection and weighting in the context of EPRV searches for terrestrial planets. 

This paper is organized as follows. Section~\ref{sec:methods} describes the methodology: selected MURaM snapshot simulations, the line lists, and the line-sensitivity analysis. Section~\ref{sec:results_sptype} presents the results across spectral type, highlighting how granulation sensitivity changes toward cooler stars and the resulting trends in line properties. Section~\ref{sec:results_depths} analyses changes in the line-forming conditions using contribution functions and links the center-of-gravity response to the granulation velocity field. Conclusions and outlook are given in Section~\ref{sec:Summary}.

\section{Methods} \label{sec:methods}
\subsection{3D model atmospheres}
We employ the three-dimensional radiative magnetohydrodynamics (MHD) code \texttt{MURaM} to simulate the near-surface convection of the Sun, G9, and K4 dwarfs using a box‐in-a‐star setup \citep{Vogler2005, Rempel2014}; see \citet{Witzke2024} for details of the numerical setup including boundary conditions. Basic parameters of  MURaM snapshots  used in this study are summarized in Table~\ref{table:muram}. The vertical extent spans sub-surface convective layers, the photosphere, and layers above the optical surface $\tau_\mathrm{Ross}=1$ where spectral lines form.  The horizontal dimensions of the G9 and K4 computational boxes were scaled relative to the solar case in proportion to the local pressure scale height. Because the characteristic horizontal size of granules scales approximately with the pressure scale height \citep{Schwarzschild1975, Magic2013}, this choice yields computational domains that naturally contain a comparable number of granules (of order $\sim$10) to the solar setup.  Magnetic fields arise self-consistently from a near-surface small-scale turbulent dynamo (SSD). All simulations assume solar composition \citep{Asplund2009}. 
\begin{table}
\centering
\caption{Summary of MURaM cubes used in the analysis.}  \label{table:muram}
\begin{tabular}{lcccccc} 
\hline
Star & $T_{\mathrm{eff}}$ (K) & $\log g$ (cm/s$^{2}$)& [Fe/H] & X$\times$Y$\times$Z size (Mm) & $N_x \times N_y \times  N_z$ size (pixels)  & $\Delta x$ (km)\\
\hline
Sun (G2)  & 5777 & 4.44 & 0.0 &  $9\times9\times5$ & $512\times 512 \times 500 $ & 17.5 \\
G9        & 5257 & 4.40 & 0.0 &  $9\times 9\times 4.8$ &  $512\times 512\times 480$ & 17.5 \\
K4        & 4593 & 4.61 & 0.0 & $4.2\times 4.2\times 2.31$ & $512\times 512\times 500$ & 8.2 \\
\hline
\end{tabular}
\end{table}
\begin{figure*}
\centering
    \includegraphics[width=1.0\textwidth]{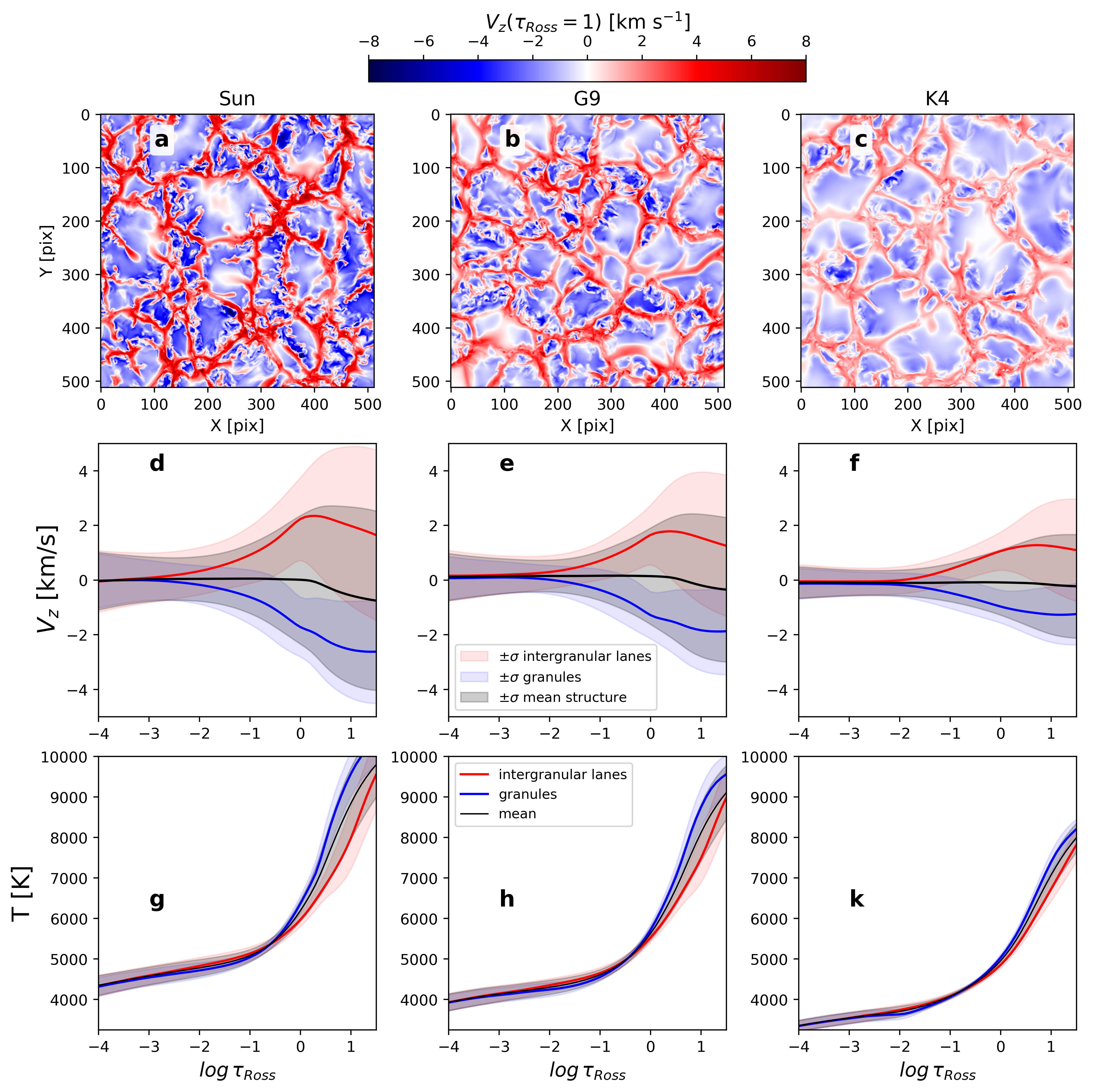}
    \caption{\textbf{MURaM simulations used in the analysis.} \textit{Top:} Maps of vertical velocity $V_{z}$ at the continuum optical-depth unity surface ($\tau_\mathrm{Ross}=1$ surface) for the Sun (left), G9 (middle), and K4 (right). We define granules and intergranular lanes by the sign of the vertical velocity $V_Z$, granules have $V_Z<0$ (up-flows, toward the observer) and intergranular lanes have $V_Z>0$ (down-flows, away from the observer).
    \textit{Middle:} Horizontally averaged $V_Z$ versus $\log\tau_\mathrm{Ross}$ for granules (blue),  intergranular lanes (red) and the entire simulated domain mean (black); shading shows the $1\sigma$ dispersion $\sigma_{V_Z}$ \footnote{
We define $\sigma_{V_z}(\tau_{\mathrm{Ross}})$ as the standard deviation of the vertical velocity $V_z$ on the $\tau_{\mathrm{Ross}}=\mathrm{const}$ surface.} of the vertical velocity.
\textit{Bottom:} Horizontally averaged temperature profiles for granules (blue), intergranular lanes (red), and the entire simulated domain mean (black) versus $\log\tau_\mathrm{Ross}$. For the panels \textit{d-k}, averages  at each $\log\tau_\mathrm{Ross}$ are computed along the corresponding constant optical depth surface. Toward cooler spectral types, the granulation pattern persists but with reduced amplitude.
}
    \label{fig:muram_models}
\end{figure*}

Figure~\ref{fig:muram_models} shows the vertical velocities and mean thermal profiles of the analyzed MURaM snapshots. Across three simulations, the vertical velocity $V_Z$ and its scatter $\sigma_{V_Z}$ reach maximum values below the optical surface ($\tau_\mathrm{Ross}=1$), reflecting the fact that convective motions strengthen below the visible surface \citep[a phenomenon often refereed to as hidden granulation, see][]{Nordlund1990, Beeck2013}. Consistent with expectations for lower $T_{\mathrm{eff}}$, the characteristic convective velocities decrease systematically from the Sun to K4 at all depths shown (both in the mean $V_Z$ and in $\sigma_{V_Z}$). This trend follows naturally from the reduced emergent radiative flux in cooler stars ($F \propto T_{\mathrm{eff}}^4$), which requires a smaller convective energy flux and therefore typically smaller convective velocities. Furthermore, because photospheric densities increase toward cooler stars, a given convective energy flux can be transported with smaller convective velocities.
The granule–intergranular lane temperature contrast likewise decreases toward cooler spectral types. The mean granule and intergranular-lane temperature stratifications intersect at $\log \tau_\mathrm{Ross}\sim-0.5$ to $-0.8$, above which the temperature contrast reverses, indicating the onset of reverse granulation \citep{Cheung2007}. 

\subsection{Spectral synthesis}
For each spectral type we analyze a single simulation snapshot (cube). An obvious way to quantify granulation sensitivity would be to compute synthetic spectra for a time series of snapshots and measure how the line diagnostics vary in time. In practice, such a time-domain approach is computationally expensive because it requires radiative transfer through many snapshots, and it is additionally complicated by coherent "box modes" (e.g., acoustic oscillations sustained by the finite simulation domain and boundary conditions) that can imprint variability distinct from the granulation signal unless one runs sufficiently long series and applies temporal filtering. These oscillations do not have the same frequency/power structure as p-modes in the real Sun, which are global standing modes. The frequency and power structure of these oscillations are affected by the finite size and boundary conditions of the local simulation box. We therefore follow the methodology of Paper~I and replace temporal variability with spatial variability within a single cube, using the ensemble of vertical columns in the snapshot as independent realizations of the surface convection pattern.
This relies on two related but distinct results validated for the solar model in Paper~I: (i) for the RV and line-sensitivity diagnostics considered here, granulation can be approximated as an ergodic process, such that spatial sampling within one snapshot is representative of temporal sampling; and (ii) a single, representative snapshot provides sufficient sampling of granulation states to yield stable line-sensitivity diagnostics. We therefore adopt the same single-cube methodology for our granulation-sensitivity analysis across spectral types.

We synthesize emergent specific intensities from each cube with the \texttt{MPS-ATLAS} radiative-transfer code in the LTE approximation, adopting a ``1.5D'' approach in which the RT is solved along many mutually parallel rays through the 3D cube while accounting for the line-of-sight velocity field \citep{Witzke2021}. For each cube and each spectral line we compute high-resolution spectra at disk center ($\mu=1$) for direct comparability to Paper~I.

\begin{figure}
\centering
    \includegraphics[width=0.5\textwidth]{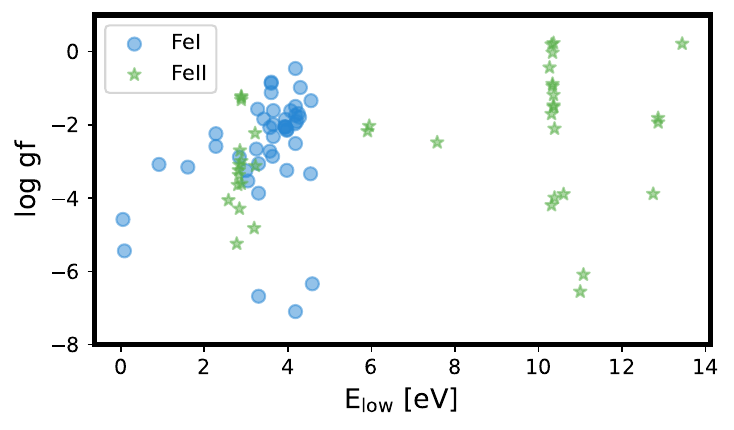}
    \caption{\textbf{Synthesized spectral lines.} Excitation potentials and oscillator strengths of analyzed Fe~{\sc i} (blue) and Fe~{\sc ii} (green) lines.}
    \label{fig:fe_real_linelist}
\end{figure}

Our line list consists of Fe~{\sc i} and Fe~{\sc ii} transitions adopted from VALD \citep{Piskunov1995,Kupka1999,Ryabchikova2015} (see see Tables B1 and B2 in Paper~I), spanning wide ranges in excitation potential and oscillator strength. Figure~\ref{fig:fe_real_linelist} shows these lines on the excitation potential - oscillator strength plane.   Spectral synthesis outputs a 3D data cube for each line, $I(x,y,\lambda)$, which we normalize by the local continuum $C(x,y)$, where $x$ and $y$ are spatial coordinates of pixels. 

Line profiles in individual simulation pixels are often complex (e.g., exhibiting multiple components due to strong radial-velocity gradients along the ray), making them inconvenient for robust line-by-line characterization. We therefore follow the approach of Paper~I and spatially average the spectra over $16\times16$-pixel sub-domains. This produces a $32\times32$ spaxel grid and yields $N_s=1024$ spatially averaged spectra per line. Paper~I showed that the derived line sensitivities are insensitive to reasonable choices of binning. We adopt $32\times32$ here for consistency.

\subsection{Line sensitivities}
For each spectral line in our line list, we measure its properties in every spaxel grid point of the cube and quantify two sensitivity parameters: (i) sensitivity of the \emph{line position} primarily to convective motions and (ii) sensitivity of the \emph{line strength} primarily to convection-driven fluctuations of the atmospheric thermal structure (while also responding to velocity-field variations).

For spaxel $i$, we compute the line center-of-gravity (COG) wavelength using the line depression $R_{i,j}=1-I_{i,j}/C_i$:
\begin{equation}
\lambda_{\mathrm{COG},i}
= \frac{\sum_{j} \lambda_j R_{i,j} \Delta\lambda_j}{\sum_{j} R_{i,j} \Delta\lambda_j}.
\label{eq:cog}
\end{equation}
Here, $i\in{1,\dots,N_s}$ indexes spaxels in the spectral cube, and $j$ is a wavelength pixel index. Therefore, $I_{i,j}$ is the emergent specific intensity in spaxel $i$ at wavelength $\lambda_j$, and $C_i$ is the local continuum intensity for that spaxel. $\Delta\lambda_j$ is the wavelength interval associated with sample $j$.
The scatter of $\lambda_{\mathrm{COG},i}$ over spaxels measures the sensitivity of the line position to convective motions,
\begin{equation}
\sigma_{\mathrm{COG}}
= \sqrt{\frac{1}{N_s}\sum_{i=1}^{N_s}\big(\lambda_{\mathrm{COG},i}-\langle \lambda_{\mathrm{COG}}\rangle\big)^2},
\end{equation}
where $\langle \lambda_{\mathrm{COG}} \rangle$ is the mean COG wavelength:
\begin{equation}
\langle \lambda_{\mathrm{COG}} \rangle = \frac{1}{N_s}\sum_{i=1}^{N_s}\lambda_{\mathrm{COG},i}.
\end{equation}
To express this in RV units, we also report the velocity-scaled quantity 
\begin{equation}
\sigma_{v}
= c\frac{\sigma_{\mathrm{COG}}}{\langle \lambda_{\mathrm{COG}} \rangle},
\label{eq:sigmav}
\end{equation}
where $c$ is the speed of light. 

The equivalent width in spaxel $i$ is
\begin{equation}
W_i = \sum_{j} R_{i,j}\, \Delta\lambda_j,
\end{equation}
and its scatter across spaxels is
\begin{equation}
\sigma_W
= \sqrt{\frac{1}{N_s}\sum_{i=1}^{N_s}\big(W_i-\langle W\rangle\big)^2},
\end{equation}
where  $\langle W\rangle$ is the spatially averaged line equivalent width:
\begin{equation}
\langle W \rangle = \frac{1}{N_s}\sum_{i=1}^{N_s} W_i.
\end{equation}
Because lines span a wide range in strength, we also use the dimensionless relative scatter
\begin{equation}
S_W = \frac{\sigma_W}{\langle W\rangle}.
\label{eq:sigmaW}
\end{equation}

Following Paper~I, we use the pair $(\sigma_v, S_W)$ to quantify sensitivity of a line to granulation. Intuitively, $\sigma_v$ traces the impact of velocity inhomogeneities on the \emph{line position} (convective shifts/asymmetries), while $S_W$ predominantly traces the impact of temperature–opacity inhomogeneities on the \emph{line strength}. However, $S_W$ should not be interpreted as a purely thermal or opacity-based diagnostic: velocity-field variations along the line of sight can also modify the equivalent width through the curve of growth by changing the local Doppler broadening and line saturation, analogous to the role of microturbulence in 1D spectrum-fitting codes. 
Thus, $\sigma_v$ and $S_W$ are complementary but not completely orthogonal measures of granulation sensitivity, and some correlation between them is expected.

As found in Paper~I, for a Sun-like star neutral lines tend to rank higher in $S_W$ and singly ionized lines higher in $\sigma_v$. Here we test how these tendencies evolve from the Sun to cooler stars.

\section{Results}

\begin{figure*}
    \includegraphics[width=\textwidth]{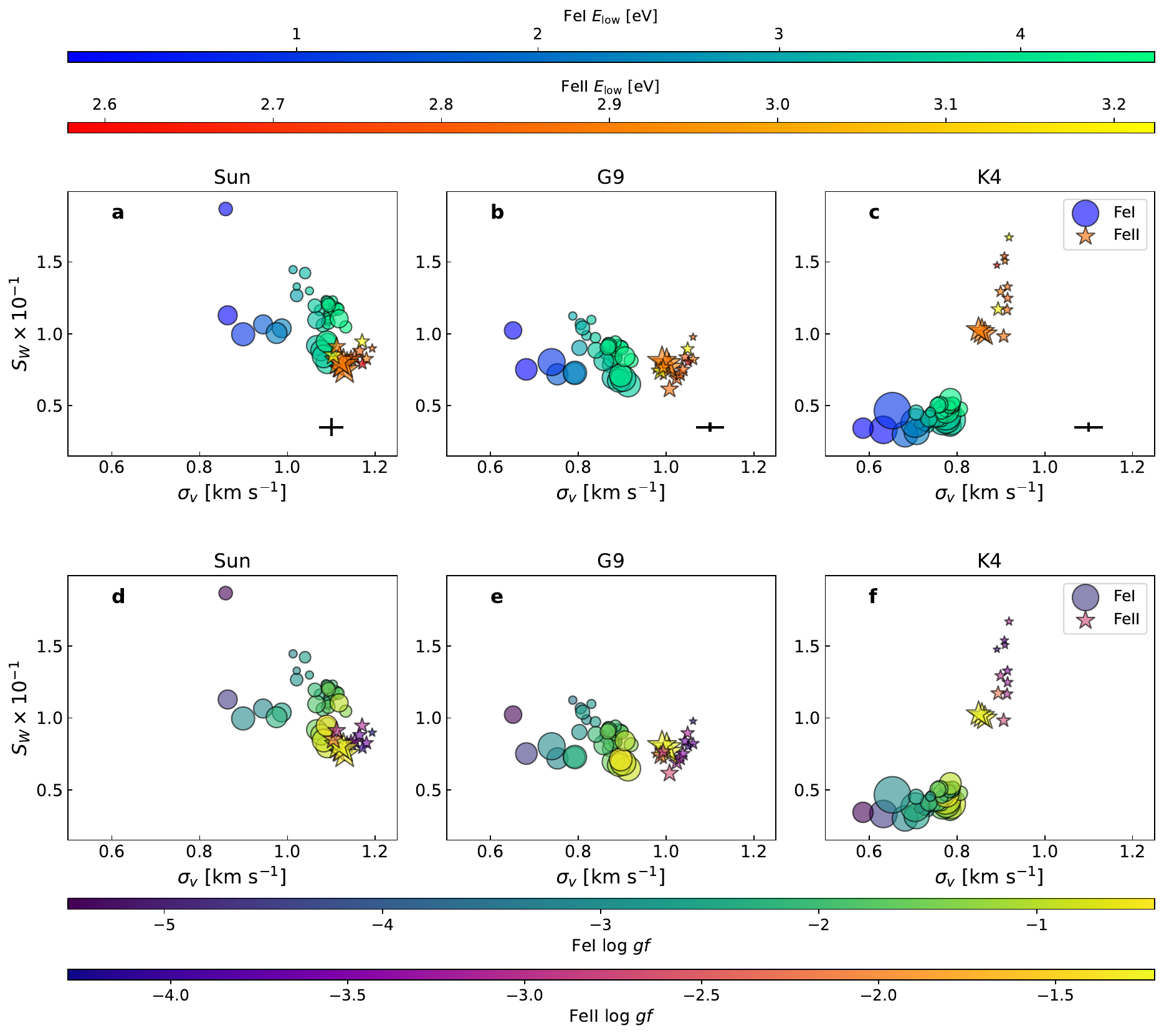}
\caption{\textbf{Dependence of Fe~{\sc i} and Fe~{\sc ii} line sensitivity to granulation on effective temperature.}
Each panel shows, for individual lines, their sensitivity to granulation as quantified by the RV-scaled line-shift scatter $\sigma_{v}$ on the $x$-axis and the fractional equivalent-width scatter $S_W$  on the $y$-axis.
Symbols denote ionization stage (circles: Fe~{\sc i}; stars: Fe~{\sc ii}) and marker sizes are proportional to the line equivalent widths.
The three columns correspond to the Sun, G9, and K4 models, respectively.
\textbf{Top row:} points are colored by 
the lower-level excitation potential $E_{\rm low}$ (shared horizontal color bars above the top row; Fe~{\sc i} and Fe~{\sc ii} shown separately).
\textbf{Bottom row:} same layout, but points are colored by oscillator strength $\log gf$ (shared horizontal color bars below the bottom row).
Black crosses in the top row indicate the characteristic cube-to-cube fluctuations in $(\sigma_v, S_W)$ estimated from five independent simulation cubes; these fluctuations reflect changes in the granule/intergranule area coverage between cubes.
We exclude extremely weak lines by requiring $W>10^{-3}$~nm to avoid numerical noise and continuum-placement uncertainty in spaxel-by-spaxel equivalent-width and COG measurements, which otherwise dominate the inferred scatter for vanishingly small line depths. 
The plotted $\sigma_v$ and $S_W$ values are local disk-center scatter metrics computed from a single simulation snapshot at $\mu=1$, not disk-integrated stellar quantities; in particular, $\sigma_v$ should not be interpreted as disk-integrated stellar RV jitter.
}
\label{fig:fe1_fe_2_strongest_lines_sensistivity}
\end{figure*}

\begin{figure*}
\centering
    \includegraphics[width=1.\textwidth]{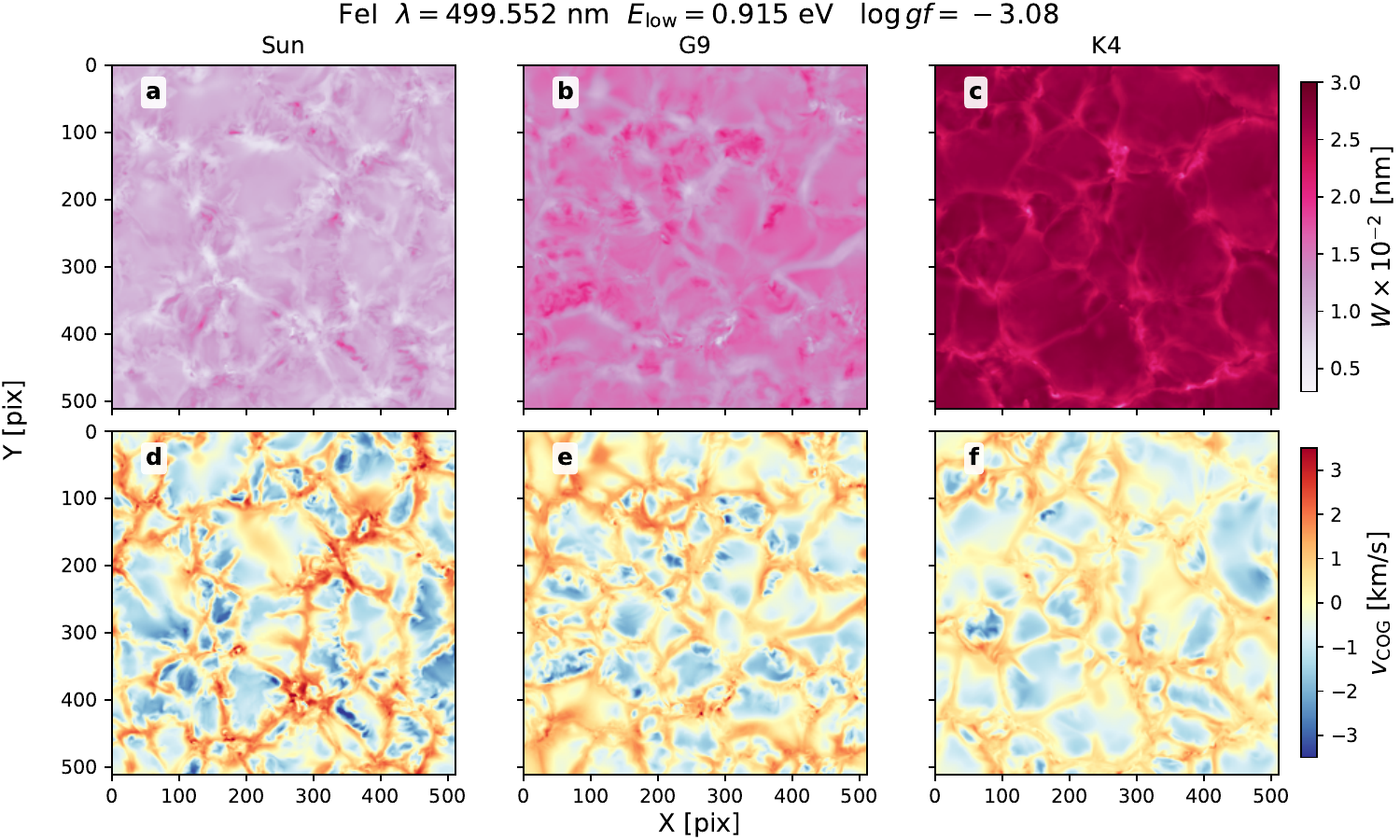}
\caption{\textbf{Spatial maps of line strength and COG velocity for a Fe~{\sc i} line at disk center.}
The maps are computed for one snapshot of each of the Sun, G9, and K4 simulations, using vertical rays at \(\mu=1\). Each pixel corresponds to one vertical column of the 3D cube. The top row shows the equivalent width, and the bottom row shows the COG velocity, for the same Fe~{\sc i} transition. The maps illustrate how the line strength and convective line shifts change toward cooler spectral type.}
\label{fig:ew_vcog_maps_fe1}
\end{figure*}

\begin{figure*}
\centering
    \includegraphics[width=1.\textwidth]{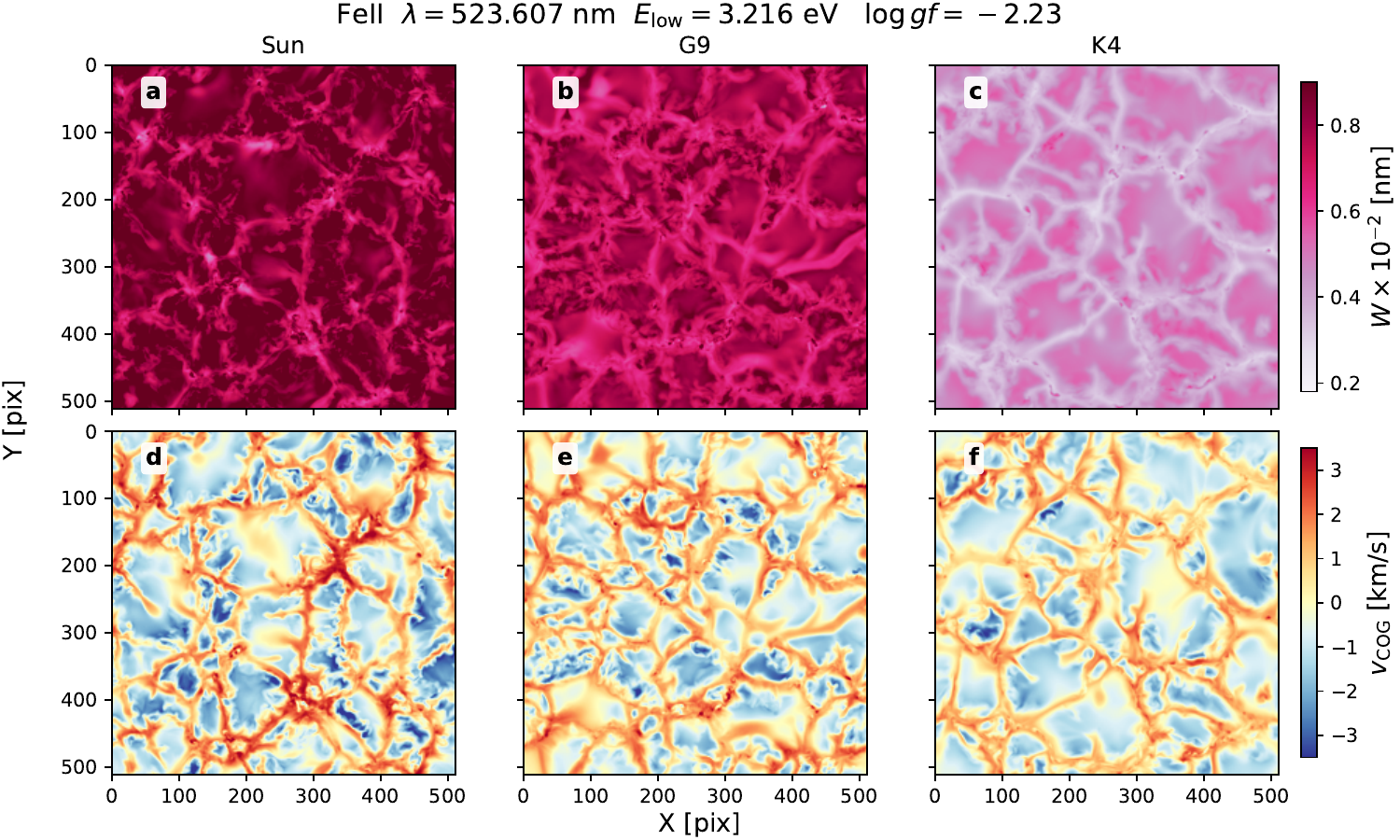}
\caption{\textbf{Spatial maps of line strength and COG velocity for a representative Fe~{\sc ii} line at disk center.}
Same as Fig.~\ref{fig:ew_vcog_maps_fe1}, but for a Fe~{\sc ii} transition.}
\label{fig:ew_vcog_maps_fe2}
\end{figure*}

\begin{figure*}
\centering
    \includegraphics[width=1.\textwidth]{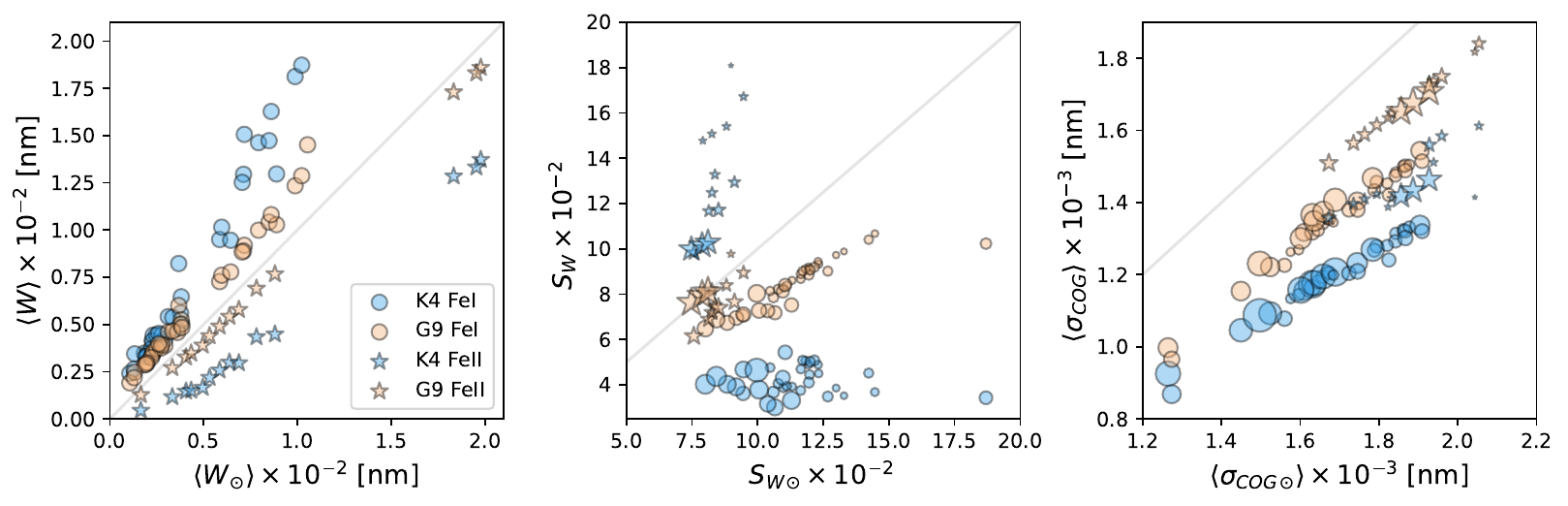}
    \caption{\textbf{Comparison of solar line sensitivities with those in the G9 and K4 models.}
      Each point compares a given line's granulation sensitivity in the G9 or K4 models (vertical axes) to its value in the solar model (horizontal axes). Symbols and colors distinguish ionization stage (Fe~{\sc i} or Fe~{\sc ii}) and spectral type, as indicated in the legend. The gray line marks equality.
      \textbf{Left:} mean equivalent width $\langle W\rangle$. 
      \textbf{Middle:} fractional equivalent‑width variability $S_W$. 
      \textbf{Right:} center‑of‑gravity wavelength scatter $\langle \sigma_{\mathrm{COG}}\rangle$.
      In the middle and right panels marker sizes are proportional to the line equivalent width $W$ (very weak lines excluded; $W>10^{-3}$~nm).
    }
\label{fig:stellar_vs_solar_ew_and_cog}
\end{figure*}

\subsection{Fe~{\sc i} vs.\ Fe~{\sc ii} granulation sensitivity from the Sun to cooler stars} \label{sec:results_sptype}
Figure~\ref{fig:fe1_fe_2_strongest_lines_sensistivity} shows our two sensitivity measures, the velocity sensitivity $\sigma_v$ and the fractional equivalent‑width scatter $S_W$, for Fe~{\sc i} and Fe~{\sc ii} lines in the Sun, G9, and K4 models.  
We note that both $S_W$ and $\sigma_v$ are computed at disk center ($\mu=1$) from a single simulation snapshot. 
In particular, $\sigma_v$ is a spatial RMS of local COG shifts within the snapshot, not a prediction of the temporal RV variability of an unresolved star.
In unresolved observations, disk integration averages the local convective velocity shifts over a large number of independent granules (for the solar case over a few millions of granules), reducing km$/$s-scale local signals to apparent granulation variability at the few to several tens of cm$/$s level.

With this definition in mind, the species trends can be interpreted as follows. Since $\sigma_v$ is the spatial RMS scatter of the per-spaxel COG shifts (Eq.~4), it  $\sigma_v$ is small when the per-spaxel COG shifts form a narrow distribution (e.g., when the line formation is effectively dominated by one granulation component). Conversely, $\sigma_v$ is larger when granules and intergranular lanes contribute comparably while imprinting distinct Doppler shifts. The solar case is identical to that presented in Paper~I and is included here as a reference for direct comparison. We therefore recover the same separation: Fe~{\sc i} lines exhibit larger $S_W$ but slightly smaller $\sigma_v$ than Fe~{\sc ii}, indicating stronger strength variability for the neutral lines but larger COG motions for the singly ionized lines.

For both Fe~{\sc i} and Fe~{\sc ii}, $\sigma_v$ decreases systematically from the Sun to K4. This is consistent with the reduction of convective velocities and granule-lane brightness contrast with decreasing $T_{\mathrm{eff}}$ (Fig.~\ref{fig:muram_models}), which reduces the net convective Doppler shift and its variability. The reduction is more pronounced for Fe~{\sc i}: by K4, the Fe~{\sc i} cluster occupies distinctly lower $\sigma_v$ than in the solar case, while Fe~{\sc ii} retains relatively large velocity sensitivity.

To visualize the spatial origin of these trends, Figs.~\ref{fig:ew_vcog_maps_fe1} and~\ref{fig:ew_vcog_maps_fe2} show maps of the equivalent width \(W\) and COG velocity \(v_{\rm COG}\) for representative Fe~{\sc i} and Fe~{\sc ii} lines in the Sun, G9, and K4 models. Each map is computed over the full horizontal extent of a single simulation snapshot, with each pixel corresponding to one vertical column of the cube. The \(v_{\rm COG}\) maps trace the granulation velocity field and show the reduced velocity contrast toward cooler models, consistent with the decrease of \(\sigma_v\). The \(W\) maps reveal the species-dependent line-strength response: the representative Fe~{\sc i} line strengthens toward K4, whereas the representative Fe~{\sc ii} line weakens, reflecting the changing ionization balance in cooler photospheres. These maps provide a visual counterpart to the \(S_W\) and \(\sigma_v\) trends measured statistically from the full line sample.

This species-dependent behavior is broadly consistent with the changing ionization balance, although for $\sigma_v$ the interpretation is likely not unique and changing line strengths probably also contribute. In the solar photosphere iron is mainly ionized, i.e.\ in the form of Fe$^+$, while neutral iron is the minority species (concentrated predominantly in cooler intergranular lanes), which provides context for the Paper~I separation recovered above: Fe~{\sc i} lines exhibit larger $S_W$ but slightly smaller $\sigma_v$ than Fe~{\sc ii} (stronger strength variability for the neutral, but larger COG excursions for the ion). As $T_{\mathrm{eff}}$ decreases, neutral iron becomes the majority species in K dwarfs, so Fe~{\sc i} opacity is expected to become less sensitive, in fractional terms, to local temperature and electron-density fluctuations. This naturally contributes to the lower and tighter $S_W$ distribution of Fe~{\sc i} in the cooler models and may also contribute to lower $\sigma_v$ by reducing the contrast between granular and intergranular contributions to the line shift. At the same time, Fe~{\sc i} lines strengthen toward cooler stars, which may further reduce their velocity sensitivity if the RV signal becomes less weighted toward the deepest, most dynamically vigorous layers.

By contrast, Fe~{\sc ii} becomes a minority species in K dwarfs, making its opacity more responsive to local temperature and electron-density variations. This naturally explains the increase and broader spread of $S_W$ for Fe~{\sc ii}$\,$ toward K4. In addition, Fe~{\sc ii} lines weaken toward cooler stars, and weaker lines in our sample tend to remain more RV-sensitive. This likely contributes to Fe~{\sc ii} retaining relatively large $\sigma_v$ even in the K4 model. Fe~{\sc ii} lines may also retain stronger differences between granular and intergranular contributions, so that the distribution of per-spaxel COG shifts remains broader. Because ionization balance, excitation potential, and equivalent width covary in the present line sample, Fig.~\ref{fig:fe1_fe_2_strongest_lines_sensistivity} does not by itself isolate which of these effects dominates the $\sigma_v$ trend. We therefore regard this interpretation of $\sigma_v$ as suggestive rather than definitive.

Consistent with this picture, with cooling Fe~{\sc i} lines strengthen while their fractional equivalent-width scatter $S_W$ decreases and tightens into a compact cluster, whereas Fe~{\sc ii} lines weaken and develop both larger $S_W$ and a broader spread. By K4, Fe~{\sc i} lines occupy the lower-left region of the $(\sigma_v, S_W)$ plane, whereas Fe~{\sc ii}$\,$ lines populate clearly higher $S_W$ and comparable or higher $\sigma_v$, preserving a practical separation between relatively ``robust'' (Fe~{\sc i}) and ``sensitive'' (Fe~{\sc ii}) line families for late-G/K dwarfs.

Within each species, the color coding in Fig.~\ref{fig:fe1_fe_2_strongest_lines_sensistivity} hints at additional structure with $\log gf$ and $E_{\rm low}$. However, these two quantities co-vary in the present line sample, so $\log gf$ should not be interpreted here as a direct proxy for line strength. Using the actual equivalent widths (marker sizes) as the more relevant strength diagnostic, lower-EW lines tend to occupy the more sensitive part of the $(\sigma_v, S_W)$ plane, whereas stronger lines cluster toward lower $S_W$; this tendency is particularly apparent for Fe~{\sc ii} at a given $T_{\mathrm{eff}}$. These patterns are plausibly related to differences in line saturation and RV-weighted formation depth, and we return to this in Sect.~\ref{sec:results_depths}.

Overall, the two-parameter plane $(\sigma_v, S_W)$ provides a compact way to quantify the regimes of line variability and to study differential behavior not only for the Sun but also for cooler late-G/K dwarfs. Fe~{\sc i} lines concentrating at low $\sigma_v$ and low $S_W$ are natural candidates for up-weighting in EPRV work, whereas Fe~{\sc ii}$\,$ lines, especially many of the lower-equivalent-width ones in the cooler models, populate the high-sensitivity region and are candidates for down-weighting or exclusion.

To quantify how equivalent widths, COG wavelength scatter, and their variability change between spectral types, we directly compare for each spectral line its sensitivities in the G9 and K4 stars to those found for the Sun in Fig.~\ref{fig:stellar_vs_solar_ew_and_cog}. For the mean equivalent width (left panel), Fe~{\sc i} and Fe~{\sc ii} move in opposite directions relative to the 1:1 line, consistent with the majority/minority switch of the two species: Fe~{\sc i} lines are generally stronger in the cooler models than in the Sun, while Fe~{\sc ii} lines are weaker. To first order, the solar values of $\langle W\rangle$ can be mapped to G9 and K4 by simple linear scalings that differ for Fe~{\sc i} and Fe~{\sc ii}. 

A similarly ordered relation is seen for the COG wavelength scatter (right panel), but with a clear species dependence: Fe~{\sc i} shifts systematically below the 1:1 relation, especially in K4, whereas Fe~{\sc ii} remains closer to it. Thus, the overall level of granulation-driven velocity noise decreases toward cooler stars, but the reduction is substantially stronger for Fe~{\sc i} than for Fe~{\sc ii}.

In contrast, the line-strength variability (middle panel) does not preserve a simple scaling. While $S_W$ in the G9 model remains moderately correlated with the solar values, the K4 points are much more dispersed, and the two species move in opposite directions: Fe~{\sc i} generally shows smaller $S_W$ than in the Sun, whereas Fe~{\sc ii} shows larger $S_W$. Thus, solar $S_W$ is a poor predictor of which lines will be strength-stable or strength-variable in the K4 model. This loss of correlation reflects the strong change in ionization balance and the associated change in the response of line opacity to local thermodynamic fluctuations. Practically, it means that line selections optimized on the Sun cannot be assumed to carry over unchanged to early-K dwarfs, particularly if they rely on equivalent-width stability rather than velocity sensitivity alone.

\subsection{Line-forming depths across spectral types} \label{sec:results_depths}
To interpret the trends in line sensitivity and to locate where in the atmosphere the RV signal is generated, we analyze line-forming depths using contribution functions with an explicit RV-information weighting. The key point is that a small Doppler shift of a spectral line $\delta v$ perturbs the continuum-normalized disk-center line intensity as $\delta f \simeq (\mathrm{d}f/\mathrm{d}v)\,\delta v$, so wavelength points with large $|\mathrm{d}f/\mathrm{d}v|$ carry most of the Doppler information. We therefore compute depth-dependent contribution functions and weight them across the line profile by $w_\lambda \propto |\mathrm{d}f/\mathrm{d}v|$ to identify the atmospheric layers that contribute most strongly to the measured RV of a given line.

For a vertical ray at disk center ($\mu=1$), the emergent monochromatic intensity follows from the formal solution of the radiative-transfer equation,
\begin{equation}
I_\lambda(\mu=1) =
\int_0^\infty S_\lambda(\tau_\lambda)\,
\mathrm{e}^{-\tau_\lambda}\,d\tau_\lambda ,
\end{equation}
where $S_\lambda$ is the monochromatic source function and $\tau_\lambda$ is the monochromatic optical depth. In LTE, the source function is given by the Planck function, i.e. $S_\lambda = B_\lambda(T)$. Changing variables from monochromatic optical depth to Rosseland optical depth using
$d\tau_\lambda = (\kappa_\lambda/\kappa_\mathrm{Ross})\,d\tau_\mathrm{Ross}$, the integrand defines the monochromatic contribution function \citep[e.g.,][Eqs.~9.3 and 9.14]{Gray2005book}. Thus, for a given wavelength $\lambda$, the contribution from a single Rosseland-optical-depth interval $\delta\tau_\mathrm{Ross}$ is
\begin{equation}
C_\lambda(\tau_\mathrm{Ross}) \delta\tau_\mathrm{Ross}  =
B_\lambda[T(\tau_\mathrm{Ross})] \,
\mathrm{e}^{-\tau_\lambda(\tau_\mathrm{Ross})}
\frac{\kappa_\lambda(\tau_\mathrm{Ross})}{\kappa_\mathrm{Ross}(\tau_\mathrm{Ross})}
\delta\tau_\mathrm{Ross} ,
\end{equation}
where $C_\lambda(\tau_\mathrm{Ross})$ is the contribution function per unit Rosseland optical depth, $\kappa_\lambda$ is the monochromatic opacity, $\kappa_\mathrm{Ross}$ is the Rosseland mean opacity, and $\tau_\mathrm{Ross}$ is the Rosseland optical depth.

We compute the contribution functions on mean 1D atmospheres obtained by horizontally averaging each 3D MURaM cube over surfaces of constant Rosseland optical depth. In this averaging, we retain not only the mean thermodynamic stratification but also the mean vertical velocity profile, \(\langle v_z\rangle(\tau_{\rm Ross})\) (see the temperature and vertical-velocity stratifications in Fig.~\ref{fig:muram_models}d--k). The resulting contribution functions are therefore computed in a plane-parallel mean atmosphere, but not in a static one: the monochromatic opacity and optical depth entering \(C_\lambda\) include the Doppler shifts associated with the depth-dependent mean vertical velocity.
We then construct the corresponding normalized cumulative contribution function
\begin{equation}
\widetilde{C_\lambda}(\tau_\mathrm{Ross}) =
\frac{\int_{0}^{\tau_\mathrm{Ross}}  C_\lambda(\tau') \delta \tau'}{\int_{0}^{+\infty}  C_\lambda(\tau')  \delta \tau'}.
\end{equation}

From $\widetilde{C_\lambda}(\tau_\mathrm{Ross})$ at every wavelength in the line profile, we extract the Rosseland optical depth $\tau_{1/2,\lambda}$ such that $\widetilde{C_\lambda}(\tau_{1/2,\lambda})=0.5$, and define the corresponding line-formation temperature  $T_{1/2,\,\lambda}$,
i.e.\ the photospheric temperature below which 50\% of the emergent disk-center intensity at that wavelength is formed \citep[following the definition from][]{Moulla2022}. Although a spectral line forms over a range of depths, $T_{1/2,\,\lambda}$ provides a single, convenient characteristic formation temperature at each wavelength point, we caution that this diagnostic is computed from quantities horizontally averaged on $\tau_\mathrm{Ross}=\mathrm{const}$ surfaces. In 3D, these surfaces are corrugated and formation depths differ substantially between granules and intergranular lanes, so $T_{1/2,\,\lambda}$ should be regarded as an effective, contribution-weighted diagnostic rather than a unique physical "formation height."

Following the approach of \cite{John2025}, for each wavelength point in the emergent line profile we compute the Doppler-sensitivity weight
\begin{equation}
w_\lambda = \left| \frac{\mathrm{d}f_\lambda}{\mathrm{d}v} \right|,
\end{equation}
where $f_\lambda$ is the continuum-normalized \emph{disk-center} intensity profile and
$v = c (\lambda - \lambda_c)/\lambda_c$ is the Doppler velocity relative to the line center $\lambda_c$.
These weights highlight the parts of the profile that respond most strongly to small velocity shifts.
We then define the flux-gradient-weighted characteristic formation temperature for each spectral line as
\begin{equation}
T_{1/2} = \frac{\sum_\lambda w_\lambda \, T_{1/2,\,\lambda}}{\sum_\lambda w_\lambda}.
\end{equation}

This RV-weighted definition biases $T_{1/2}$ toward the formation depths from which most of the RV signal originates. Specifically, $\left|\mathrm{d}f_\lambda/\mathrm{d}v\right|$ is largest where the profile is steepest (the line flanks), which typically occurs in the line wings rather than in the line core. Because the wings form closer to the continuum-formation layers than the line core, they typically form deeper in the photosphere and therefore sample higher temperatures. Moreover, in the   400-450~nm wavelength range, where the considered lines are located,  the continuum opacity is typically lower than at longer visible wavelengths, so the continuum layer is shifted to deeper hotter layers. In addition, because we consider disk center intensities ($\mu=1$), the emergent intensity samples deeper layers than disk-integrated flux, further shifting the characteristic formation temperature to hotter values.
A useful way to understand these trends is via the Eddington–Barbier relation, which links $I_\lambda(\mu=1)$ to conditions near $\tau_\lambda \approx 1$ \citep[e.g.,][Eq.~2.44]{Rutten2003}. As a result, the RV-weighted $T_{1/2}$ should be interpreted as the temperature of the photospheric layers that dominate the Doppler information content of the line (primarily the line flanks), rather than the cooler heights associated with the line core.

\begin{figure*}[htb!]
\centering
    \includegraphics[width=1.0\textwidth]{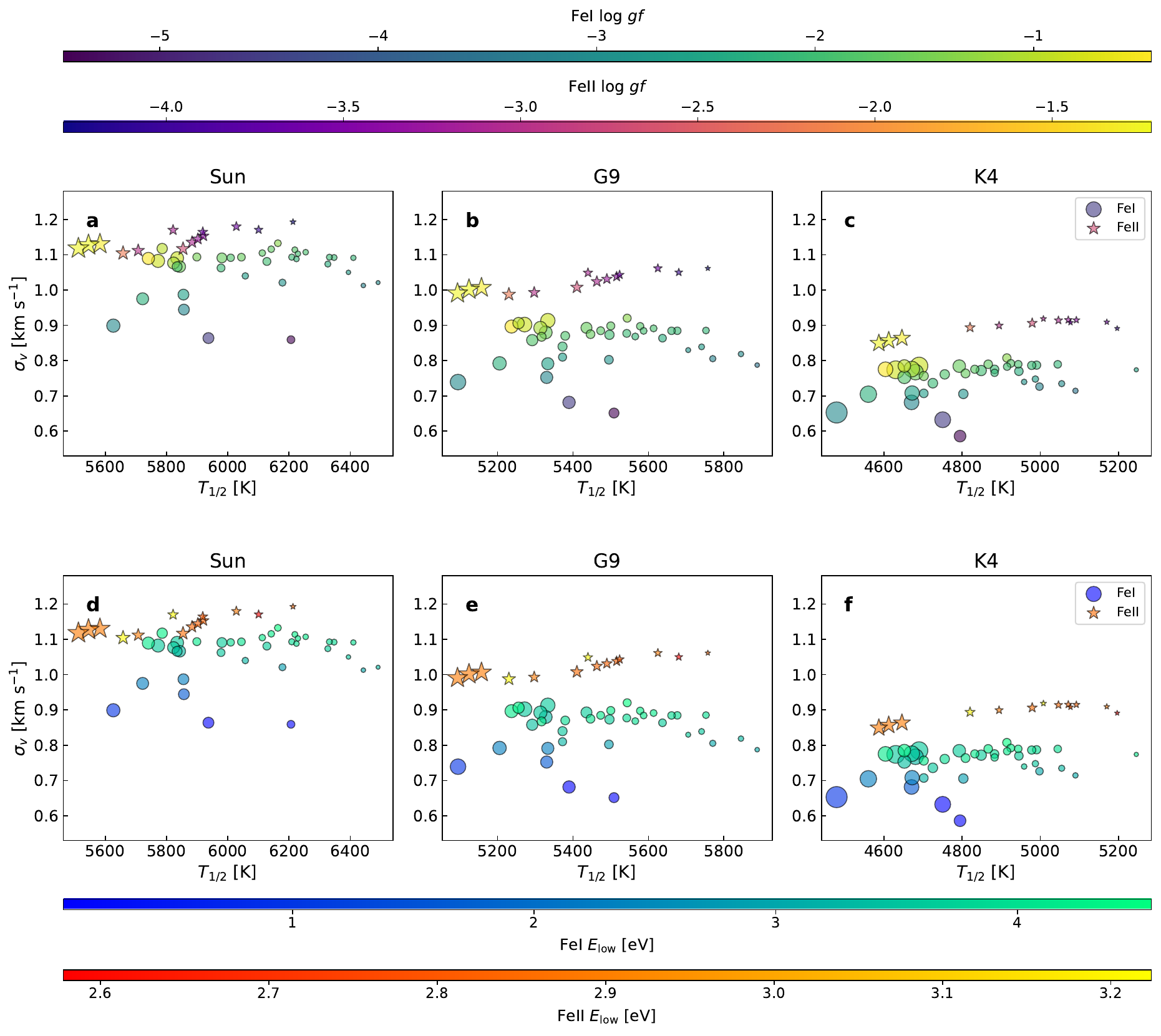}
\caption{\textbf{Standard deviation of the line center of gravity as a function of RV-weighted line-formation temperature.}
Each panel shows $\sigma_{\mathrm{COG}}$ as a function of the RV-weighted formation temperature $T_{1/2}$, defined from the cumulative \emph{disk-center} ($\mu=1$) contribution functions and weighted across the line profile by $w_\lambda \propto \left|\mathrm{d}f_\lambda/\mathrm{d}v\right|$ (see Section~\ref{sec:results_depths}). Symbols denote species (circles: Fe~{\sc i}; stars: Fe~{\sc ii}). The three columns correspond to the Sun, G9, and K4 models, respectively. \textbf{Top row:} points are colored by the lower-level excitation potential $E_{\rm low}$ (shared horizontal color bars above the top row; Fe~{\sc i} and Fe~{\sc ii} shown separately). \textbf{Bottom row:} same layout, but points are colored by oscillator strength $\log gf$. Marker sizes are proportional to the line equivalent width $W$ (very weak lines excluded; $W>10^{-3}$~nm). Because the RV-weighting emphasizes the line flanks and we use disk-center intensities in the blue (400-450\,nm), $T_{1/2}$ is weighted toward flank-forming layers, i.e. photospheric depths intermediate between the line core and the continuum, rather than toward the highest, coolest core-forming layers. As in Fig. 3, \(\sigma_v\) values are local disk-center scatter metrics computed from a single simulation snapshot at $\mu=1$, not disk-integrated stellar quantities.
}
\label{fig:line_formation_and_sens}
\end{figure*}

Figure~\ref{fig:line_formation_and_sens} shows the standard deviation of the line center of gravity as a function of $T_{1/2}$ for Fe~{\sc i} and Fe~{\sc ii} lines in the Sun, G9, and K4 models. 
For Fe~{\sc ii} lines, $\sigma_{v}$ increases systematically toward higher $T_{1/2}$ in all three models: weak lines (small equivalent widths and lower oscillator strengths) forming deeper (at hotter layers) are more strongly affected by granulation because they sample regions where the vertical convective velocities are larger, and the velocity field imprints stronger Doppler asymmetries on the emergent profiles.  Fe~{\sc i} lines do not follow a comparably simple monotonic sequence with $T_{1/2}$.  Instead, their distribution is structured by both excitation potential and equivalent width.
The lowest-excitation Fe~{\sc i} lines cluster at relatively low $T_{1/2}$ and low $\sigma_v$; importantly, these are mostly strong lines. Conversely, within the higher-excitation subset (e.g.\ $E_\mathrm{low} > 2$~eV), smaller-equivalent-width lines tend to reach higher $T_{1/2}$, consistent with weaker lines sampling deeper, hotter layers. Part of the vertical scatter at a given $T_{1/2}$ tracks excitation potential, with higher-excitation Fe~{\sc i} lines generally showing larger $\sigma_v$ than the low-excitation subset. This indicates that the convective shift variability depends not only on the RV-weighted formation depth, but also on line excitation and line strength, which affect how strongly granular and intergranular contributions enter the COG signal. For observers, this means that line strength is not merely a selection criterion for detectability: even among lines with similar excitation potential, choosing weaker versus stronger lines changes which atmospheric depths dominate the RV information and can therefore change the line’s granulation-driven RV scatter. The overall level of $\sigma_{v}$ decreases from the Sun to the cooler K-dwarf model, consistent with the reduced convective velocity amplitudes and granule-integranular lane intensity contrasts in the cooler simulations (Fig.~\ref{fig:muram_models}).

Overall, this figure shows that the impact of granulation on spectroscopic RVs is primarily set by the depth at which a line contributes most of its RV information, with Fe~{\sc ii} providing a comparably much cleaner mapping between $T_{1/2}$ and $\sigma_\mathrm{COG}$. The weak downturn of $\sigma_v$ at the highest $T_{1/2}$ for Fe~{\sc ii} should be interpreted cautiously, since this part of the sample is represented by only a small number of weak, low-$gf$ lines rather than by an independent sequence in formation temperature alone. Since $\sigma_v$ is the spatial RMS of the local COG shifts, changes in line strength, excitation potential, and oscillator strength can modify the relative weighting of granular upflows and intergranular downflows, thereby narrowing the distribution of local COG shifts even for lines with relatively high $T_{1/2}$.

\section{Summary and outlook}
\label{sec:Summary}
We extended the granulation-sensitivity framework introduced in Paper~I from the Sun to cooler main-sequence dwarfs (G9 and K4) using Fe~{\sc i} and Fe~{\sc ii} lines. Moving from solar to K-dwarf effective temperatures, near-surface convection becomes less vigorous and the Fe ionization balance shifts, with Fe~{\sc i} becoming dominant and Fe~{\sc ii} becoming comparatively rare. As a result, neutral and ionized lines separate more cleanly in their response to granulation: Fe~{\sc i} lines show smaller fractional variations in line strength and smaller line-shift scatter, while Fe~{\sc ii} lines remain comparatively sensitive and can exhibit comparatively large fractional variability, even as the overall convective RV variability decreases toward K4.

We link these trends to where the RV information is generated in the atmosphere. A simple message emerges: lines whose RV information is dominated by deeper photospheric layers tend to be more affected by granulation, because those layers contain larger vertical velocity fluctuations. 
In the present sample, higher-excitation lines often correspond to deeper, hotter RV-weighted formation and therefore larger granulation-driven line-shift variability. In addition, within subsets of similar excitation potential, lower-equivalent-width lines tend to form deeper and to show larger velocity sensitivity. Here line strength should be understood in terms of the actual equivalent width $W$, not $\log gf$, since $\log gf$ and $E_{\rm low}$ co-vary in our sample. This depth-based ordering persists from the Sun to G9 and K4.  

For observers, the practical implication is that \emph{stellar-type-specific line selection is necessary}: line lists or spectral masks optimized on the Sun should not be assumed to transfer unchanged to late-G and early-K dwarfs. In particular, selections based primarily on equivalent-width stability can fail to minimize RV scatter because the mapping between line strength, formation depth, and velocity sensitivity changes with spectral type. In our models, many Fe~{\sc i} lines provide robust candidates for up-weighting (low scatter in both line shift and line strength), whereas Fe~{\sc ii} lines, especially weak transitions, are prime candidates for down-weighting or exclusion in EPRV masks. 

In parallel, many analyses model stellar variability at the level of the extracted RV/cross-correlation function time series using pre-whitening, Gaussian-process regression, and other data-driven approaches (e.g., \citealt{Cameron2021}). Because such methods operate on a line-averaged observable, they are therefore complementary to the strategy pursued here: our results enable spectral-type-specific line selection and  weighting upstream of RV extraction.

The present analysis is restricted to disk-center intensities. Related center-to-limb effects on granulation sensitivity were discussed for the solar case in Paper I. Future work should extend the disk-center formation-temperature analysis to disk-integrated spectra for late-G/K models using an appropriate flux contribution-function formalism \citep[e.g.][]{Palumbo2025}. Furthermore, we aim to broaden the granulation-sensitivity analysis across a wider grid of stellar parameters and additional species, and ultimately implement and test spectral-type-specific masks on observed high-cadence RV time series.

\begin{acknowledgments}
VV and NK acknowledge support from the Max Planck Society under the grant ``PLATO Science'' and from the German Aerospace Center under ``PLATO Data Center'' grant   50OO1501.
KS, AIS, DV, VW, and ACC acknowledge support from ERC Synergy Grant REVEAL under the European Union’s Horizon 2020 research and innovation program (grant no. 101118581). 
TB and AIS acknowledge support from DFG grant SH1489/1. L.G. acknowledges partial funding from the DLR grant ``PLATO Data Center'' (FKZ 50OO1501 and 50OP1902). We gratefully acknowledge the computational resources provided by the  Raven supercomputer systems of the Max Planck Computing and Data Facility (MPCDF) in Garching, Germany.
\end{acknowledgments}

\software{
matplotlib \citep{matplotlib}, 
numpy \citep{numpy}, 
scipy \citep{scipy}
}

\bibliography{bilbiography}{}
\bibliographystyle{aasjournalv7}
\end{document}